\documentclass[12pt]{article}
\usepackage[esperanto]{babel}
\usepackage{epsfig}

\newcommand{\bea}{\begin{eqnarray}}
\newcommand{\eea}{\end{eqnarray}}
\newcommand{\dd}{\mathrm{d}}

\begin{document}
\selectlanguage{esperanto}

\title{\bf Relativeca Dopplera efekto ^ce unuforme akcelata movo -- I}  
\author{F.M. Paiva \\ 
{\small Departamento de F\'\i sica, U.E. Humait\'a II, Col\'egio Pedro II} \\
{\small Rua Humait\'a 80, 22261-040  Rio de Janeiro-RJ, Brasil; fmpaiva@cbpf.br} 
\vspace{.7ex} \\
A.F.F. Teixeira \\
{\small Centro Brasileiro de Pesquisas F\'\i sicas} \\
{\small 22290-180 Rio de Janeiro-RJ, Brasil; teixeira@cbpf.br}} 

\maketitle 

\begin{abstract} 
Observanto ^ce rekta movo ^ce konstanta propra akcelo pasas preter restanta fonto de unukolora radiado. ^Ce la special-relativeco, ni priskribas la observatan Doppleran efekton. Ni anka^u priskribas la  interesan nekontinuan efekton se trapaso okazas.   
 \\ - - - - - - - - - - - \\ 
An observer, in rectilinear motion under constant proper acceleration, passes near a source of monochromatic radiation at rest. In the context of special relativity, we describe the observed Doppler effect. We describe also the interesting discontinuous effect when riding through occurs. An English version of this article is available.   
\end{abstract}

\section{Enkonduko} \label{Unua}
^Ce special-relativeco, la rekta movo ^ce konstanta propra akcelo estis studata, ekzemple, per M\o ller \cite[pa^go 72]{Moller}, per Rindler \cite[pa^go 49]{Rindler}, per Dwayne Hamilton \cite{Hamilton} kaj per Landau kaj Lifshitz \cite[pa^go 22]{LL}. Anka^u bone konata estas la Dopplera efekto de unukolora radiado eligita el restanta fonto, vidata per observanto tiel akcelata. Tamen, oni ofte evitas la studon de okazoj kies la observanto pasas preter la eliganto. ^Ciokaze estas granda ^san^go de la observata frekvenco. ^Sajnas ke nur Cochran \cite{Cochran} apena^u menciis tion.  

^Ci tie ni da^urigas Rothenstein, Popescu, kaj Gruber \cite{RothPop}, kiu fre^sdate studis okazojn kies la observanto alproksimi^gas al la fonto, sed ne pasas preter ^gi. Kontra^ue, nia studo konsideras preterpasojn, kaj anka^u trapasojn. 

\section{Kinematika priskribo} \label{Dua}
\begin{figure}[t]                                             \centerline{\epsfig{file=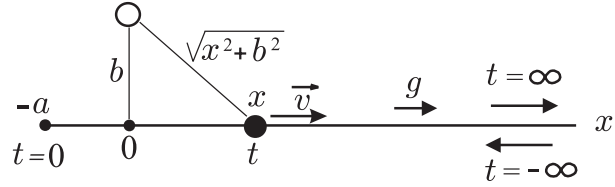,width=8cm}} 
\caption{Luma fonto (blanka sfereto) estas fiksa distance $b$ de akso $x$. Observanto (nigra sfereto) venis el $x\!=\!\infty$ ($t\!=\!-\infty$) ^gis $x\!=\!-a$ ($t\!=\!0$). Poste li revenos al $x\!=\!\infty$ ($t\!=\!\infty$).}                                        \label{Figuro1} 
\end{figure}

Luma fonto estas fiksa distance $b$ de akso $x$, kaj da^ure eligas  radiadon kun konstanta frekvenco $\nu$. Kiel figuro~\ref{Figuro1} montras, observanto venis el $x\!=\!\infty$, kiam\footnote{Por mezuri $t$ ^ce la inercia referenco-sistemo de la fonto, horlo^garo estas supozata distribuata la^u akso $x$. ^Ciu horlo^go estas sinkrona al tio de la fonto.} $t\!=\!-\infty$.  Lia komenca rapido estis multe granda, $v\!\approx\!-c$, sed li grade bremsi^gis ka^uze de lia propra konstanta malakcelo $g\!>\!0$. La observanto apena^u atingis $x\!=\!-a\!<\!0$       ^ce momento $t\!=\!0$, kaj tuj ekrevenos same akcelate al $x\!=\!\infty$. Ni studos la variadon de la {\em vidata}\footnote{Ni emfazas la verbon ``vidi'' por signifi ke tio ne estas nur kalkulita koordinata efekto, sed vere estas la {\em observata} efekto.} frekvenco $\nu_{obs}$ dum la movado ekde $t\!=\!-\infty$ ^gis $t\!=\!\infty$.  

Kontra^ue kutima literaturo, kie $a\!=\!-c^2/g$, atentu ke ^ci tie $a$ estas iom ajn pozitiva. Tiamaniere, ni ebligas la interesajn studojn de preterpaso ($b\!\neq\!0$) kaj trapaso ($b\!=\!0$).  

Por simpligi formulojn ni formale konsideros $c\!=\!1$ kaj $g\!=\!1$. Por aperigi la arbitrajn valorojn sufi^cas substitui  
\bea                                                      \label{reak}
a\!\rightarrow\! ag/c^2\ , \hskip2mm b\!\rightarrow\!bg/c^2\ , \hskip2mm x\!\rightarrow\!gx/c^2\ , \hskip2mm v\!\rightarrow\!v/c\ , \hskip2mm t\rightarrow\!gt/c\ , \hskip2mm \tau\!\rightarrow\!g\tau\!/c\ . 
\eea  

^Ce la inercia referenco-sistemo de la fonto, la diferenciala ekvacio de movado de observanto ^ce propra akcelo $g$ estas~\cite[pa^go 49]{Rindler}\footnote{Uzante (\ref{reak}), la unua de (\ref{eq}) skribi^gus ${\rm d}\!\left((v/c)/\sqrt{1-(v/c)^2}\,\right)/{\rm d}(gt/c)=1$\ , kiu simpli^gus al la pli familiara ${\rm d}\!\left(v/\sqrt{1-v^2/c^2}\,\right)\!/{\rm d}t=g$\ .}
\bea                                                      \label{eq}
\frac{\dd}{\dd t}\left(\frac{v}{\sqrt{1-v^2}}\right)=1\ , 
\hskip3mm 
v:=\frac{\dd x}{\dd t}\ .  
\eea
Tiu, integralante kun $v\!=\!0$ kiam $t\!=\!0$, estante $g$ konstanta, esti^gas la observanta rapido 
\bea                                                      \label{vt}
v=\frac{t}{\sqrt{1+t^2}}\ ; 
\eea 
atentu ke la signumoj de $v$ kaj $t$ estas samaj. Tiu, integralante kun $x\!=\!-a$ kiam $t\!=\!0$, esti^gas la spaco-tempa hiperbolo $(x+1+a)^2-t^2\!=\!1$\ ; do la pozicio de observanto ^ce momento $t$ estas
\bea                                                      \label{xt}
x=-a+\sqrt{1+t^2}-1\ .  
\eea

\begin{figure}[ht]                                             
\centerline{\epsfig{file=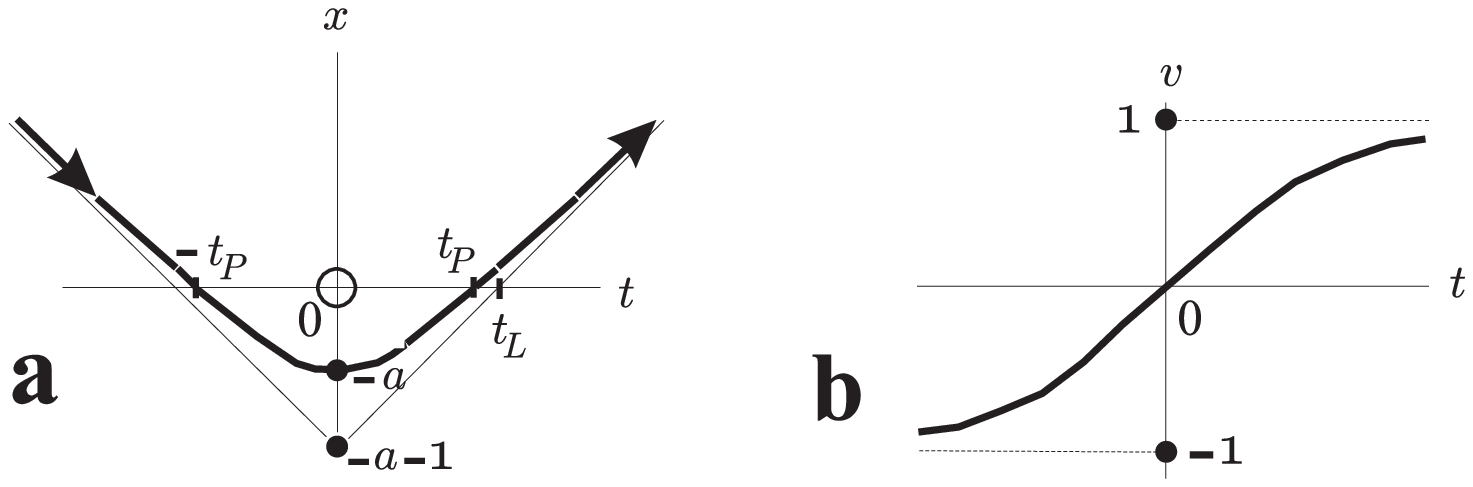,width=12cm}} 
\caption{Funkcioj $x(t)$ kaj $v(t)$.
\newline 
{\bf 2.a} La radianta fonto (blanka sfereto) estas fiksa en $x\!=\!0$, kaj la observanto estas ^ce hiperbola movado (\ref{xt}); li venis el $x\!=\!\infty$, pasas preter la fonto kiam $t\!=\!-t_P$, atingas $x\!=\!-a\!<\!0$ kiam $t\!=\!0$, denove pasas preter la fonto kiam $t\!=\!t_P$, kaj fine revenos al $x\!=\!\infty$.
\newline
{\bf 2.b} La observanta rapido (\ref{vt}), kiu kontinue varias de $-c$ (kiam $t\!=\!-\infty$) ^gis $c$ (kiam $t\!=\!\infty$).} \label{Figuro2}
\end{figure}

La preter(tra)paso okazas dufoje ^ce $x\!=\!0$, kiel figuro~\ref{Figuro2}.a montras; la du momentoj de preter(tra)paso estas $t\!=\!\mp\,t_P$, estante  
\bea                                                      \label{tP}
t_P:=\sqrt{a^2+2a}\ . 
\eea 
Figuro~\ref{Figuro2}.a montras anka^u, ke lumo atinganta observanton devas eli^gi je $t\!<\!t_L$, estante 
\bea                                                      \label{tL}
t_L:=a+1\ ;  
\eea
atentu ke $t_L^2\!=\!t_P^2+1$\ , kaj ke nek $t_P$ nek $t_L$ dependas de $b$\ .

\section{Dopplera efekto kun preterpaso}
\begin{figure}[ht]                                             
\centerline{\epsfig{file=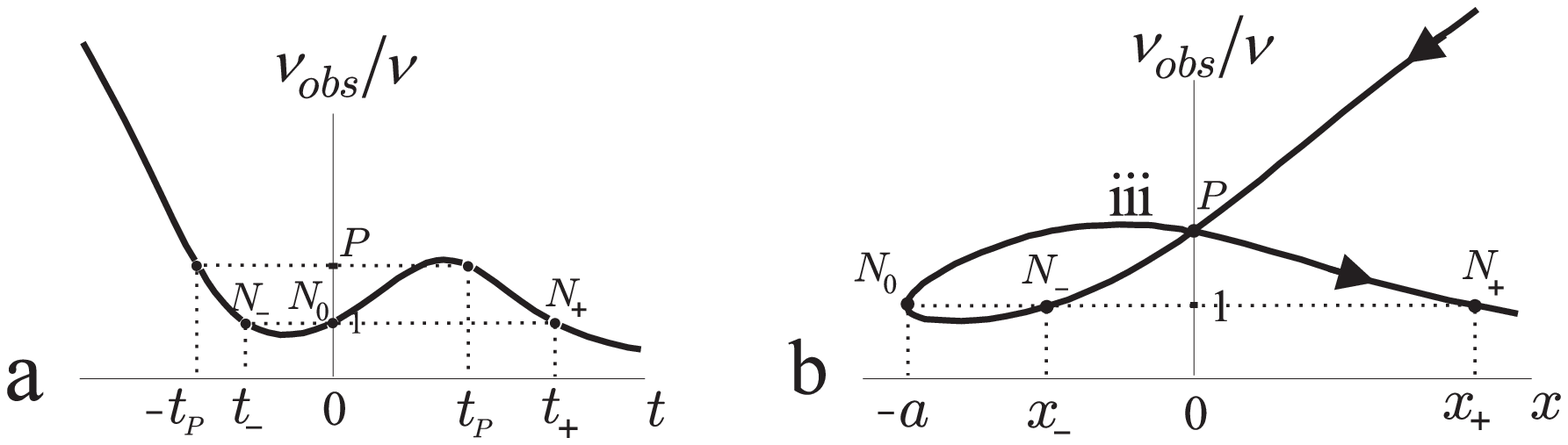,width=13cm}} 
\caption{
\newline 
{\bf \ref{Figuro3}.a} $\nu_{obs}/\nu$ kiel funkcio de fonta horlo^garo  $t$.   
\newline  
{\bf \ref{Figuro3}.b} $\nu_{obs}/\nu$ kiel funkcio de observanta pozicio $x$. Sagoj montras la pozitivan tempan fluon. Se $b$ estas sufi^ce granda ($b\!>\!c\,t_P$), la inklinacio de {\bf iii} ^ce $P$ estas pozitiva.} 
                                                          \label{Figuro3}
\end{figure}

Supozu ke movi^ganta observanto ricevas du lumajn signalojn, je momentoj $t$ kaj $t+\dd t$, ^ce pozicioj $x$ kaj $x+\dd x$. Tiuj signaloj estis eligataj je la anta^uaj momentoj $t_E$ kaj $t_E+\dd t_E$. Figuro~\ref{Figuro1} montras ke  
\bea                                                      \label{tE}
t_E\!=\!t-\sqrt{x^2+b^2}\ .
\eea 
^Ce la horlo^go de movi^ganta observanto, la {\em propra}\,tempo inter la du ricevoj estas $\dd\tau\!=\!\sqrt{1-v^2}\dd t$, estante $v(t)\!=\!\dd x/\dd t$ la rapido de la observanto. La frekvenco $\nu_{obs}$ {\em vidata} per observanto rilatas, al la frekvenco $\nu$ de la lumo eligata, kiel $\nu_{obs}/\nu\!=\!\dd t_E/\dd\tau$\ . Simplaj kalkuloj donas 
\bea                                                 \label{semnome}
\nu_{obs}/\nu\!=\!\frac{1}{\sqrt{1-v^2}}\left(1-\frac{xv}{\sqrt{x^2+b^2}}
\right)\ ,
\eea 
kiu uzante (\ref{vt}) kaj (\ref{xt}) skribi^gas  
\bea                                                     \label{Dop}
\nu_{obs}/\nu\!=\!\sqrt{1+t^2}-\frac{t\,x}{\sqrt{x^2+b^2}}\ .
\eea 
Atentu ke tiuj $t$ kaj $x$ estas la momento kaj pozicio de {\em ricevo} de radiado. Por havi $\nu_{obs}/\nu$ kiel funkcio de nur $t$, uzu (\ref{xt}); kaj por havi $\nu_{obs}/\nu$ kiel funkcio de nur $x$, uzu la inverso de (\ref{xt}),
\bea                                                       \label{t}
t\!=\!\epsilon\sqrt{(a+1+x)^2-1}\ , \hskip3mm \epsilon\!:=\!t/|t|\ .
\eea 
Figuro~\ref{Figuro3} montras tiujn du funkciojn. Memoru ke se $\nu_{obs}/\nu\!>\!1$, okazas viol-delokigon; se $\nu_{obs}/\nu\!<\!1$, okazas ru^g-delokigon; kaj se $\nu_{obs}/\nu\!=\!1$, la Dopplera efekto estas nula.  

Tri rimarkoj ^sajnas gravaj pri tiuj grafikoj. Unue, ^ce figuro~\ref{Figuro3}.a vidu asimptoton $\nu_{obs}/\nu\!=\!-2t$ kiam  $t\!\rightarrow-\infty$, kaj hiperbolon $\nu_{obs}/\nu\!=\!(1+b^2)/(2t)$ kiam $t\!\rightarrow\infty$. Kaj ^ce figuro~\ref{Figuro3}.b la asimptoto estas $2(x+a+1)$ kiam $t\!\rightarrow-\infty$\ , kaj la hiperbolo estas $(1+b^2)/(2x)$ kiam $t\!\rightarrow\!\infty$.

Due, ni kalkulas la Doppleran efekton ^ce la momentoj de preterpaso (ortan Doppleran efekton). Uzante $\{t\!=\!\mp\,t_P, x\!=\!0\}$ kaj (\ref{tP}) ^ce (\ref{Dop}), la efekto esti^gas $1+a$, kiun ni nomas $P$: 
\bea                                                       \label{D0}
P:=\!1+a\ .
\eea 
Atentu ke ^gi ne dependas de $b$, kaj ke la orta Dopplera efekto pri  preterpasanta observanto estas {\em viol}-delokigo. Estas interesa noti ke pri preterpasanta fonto kaj restanta observanto, la orta Dopplera efekto estas {\em ru^g}-delokigo, kion ni vidos ^ce estonta artikolo. 

Trie, ni kalkulas la momentojn kaj poziciojn de nula Dopplera efekto,   farante $\nu_{obs}/\nu\!=\!1$ ^ce (\ref{Dop}), kun $b\!\neq\!0$. Tiu okazas trifoje, kiel figuro~\ref{Figuro3} montras. Krom la bonkonata $N_0\{t\!=\!0, x\!=\!-a\}$ de momente haltata observanto $(v\!=\!0)$, anka^u okazas $N_-$ kaj $N_+$, en kiuj la observanto estas movi^ganta \cite{PaivaTeixeira2006}. La du novaj pozicioj\,, momentoj\,, kaj rapidoj estas (memoru, $\epsilon\!=\!\mp1$ estas la signumo de $t$) 
\bea                                                      \label{xtv}
x_\epsilon\!=\!\frac{b^2}{4}\left(1+\epsilon\sqrt{1+8a/b^2}\right)\ , \hskip3mm 
t_{\epsilon}\!=\!\epsilon\sqrt{(a+1+x_\epsilon)^2-1}\ , \hskip3mm v_\epsilon=\frac{t_\epsilon}{a+1+x_\epsilon}\ . 
\eea 
Atentu ke $x_+\!>\!|x_-|$\ , $t_+\!>|t_-|$\ , kaj $v_+\!>\!|v_-|$\ .

Nun ni resumas tiujn rezultojn. La observanto unue {\em vidas} malgrandi^gantan viol-delokigon ^gis $\{t_-,x_-\}$. Je $t_-$ estas nula Dopplera efekto. De $t_-$ ^gis $\{t\!=\!0,x\!=\!-a\}$ li {\em vidas} grandi^gantan kaj poste malgrandi^gantan ru^g-delokigon. Je $\{t\!=\!0,x\!=\!-a\}$ li estas momente haltata kaj do {\em vidas} nulan Doppleran efekton. De $\{t\!=\!0,x\!=\!-a\}$ ^gis $\{t_+,x_+\}$ li {\em vidas} viol-delokigon, kiu grandi^gas kaj poste malgrandi^gas. En $\{t_+,x_+\}$ li {\em vidas} nulan efekton. Poste li {\em vidas} ru^g-delokigon kiu grandi^gas. Atentu ke ^ce la momentoj $\mp t_P$ de preterpaso ($b\!\neq\!0$) ne okazas speciala ^san^go de Dopplera efekto.  

\section{Dopplera efekto kun trapaso} 

Se $b\!=\!0$ trapaso okazas ^ce $x\!=\!0$,  kaj pluraj variabloj malkontinui^gas je la du momentoj $\mp t_P$ de trapaso. Fakte, la (\ref{Dop}) kun $b\!=\!0$ simpli^gas al    
\bea                                                   \label{Dop0}
\nu_{obs}/\nu\!=\!\sqrt{1+t^2}-\epsilon_x t \ ,
\eea
estante $\epsilon_x\!:=\!x/|x|$ la signumo de $x$. Uzante ekvaciojn (\ref{xt}) kaj (\ref{tP}) ni konkludas ke la signumo de $x$ estas sama al la signumo de $t^2-t_P^2$, tial (\ref{Dop0}) estas skribebla kiel funkcio de nur $t$, 
\bea                                                 \label{Dnova1}
\nu_{obs}/\nu\! = 
\left\{\begin{array}{l}
\sqrt{1+t^2}-t, \hskip3mm |t|\!>\!t_P\ , 
\\
\sqrt{1+t^2}+t, \hskip3mm |t|\!<\!t_P\ . 
\end{array}\right.
\eea 

\begin{figure}[ht]                                             
\centerline{\epsfig{file=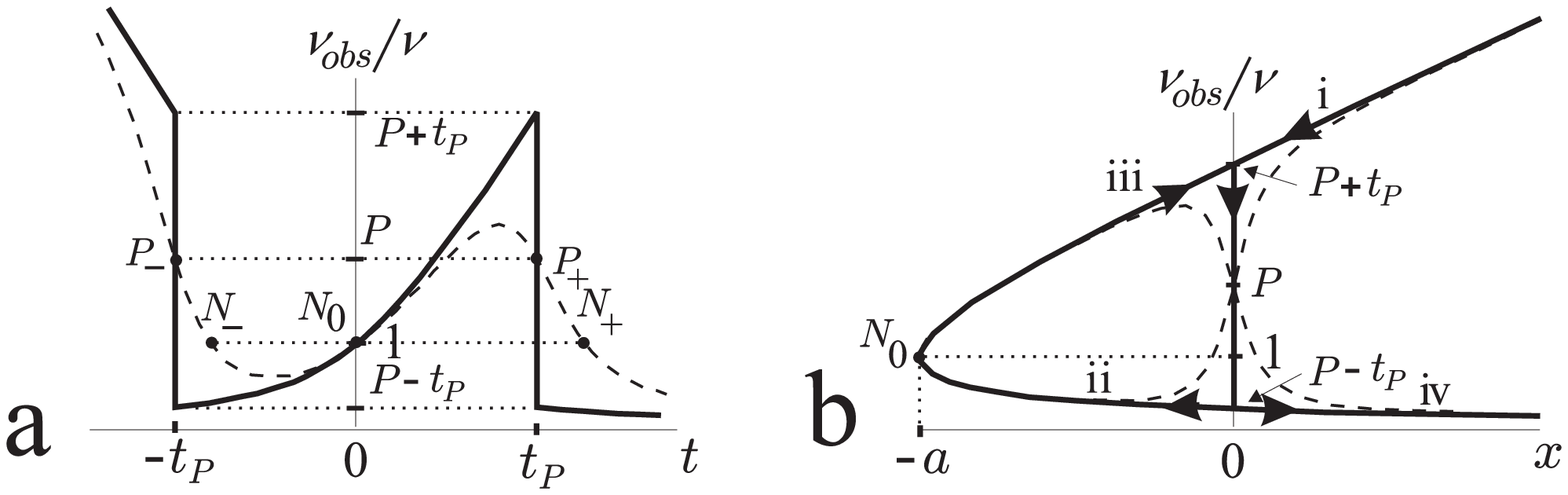,width=14cm}} 
\caption{
\newline 
{\bf \ref{Figuro4}.a} montras $\nu_{obs}/\nu$ kiel funkcio de horlo^gara tempo $t$. La plenlinio rilatas al trapaso ($b\!=\!0$, ekvacioj (\ref{Dnova1})). Por komparado, la streklinio rilatas al  preterpaso (\ref{Dop}), kun $b\!=\!a/2$.  
\newline  
{\bf \ref{Figuro4}.b} montras $\nu_{obs}/\nu$ kiel funkcio de observanta pozicio $x$. La plenlinio rilatas al trapaso ($b\!=\!0$, ekvacio  (\ref{Dop0x})); atentu la tempan ordon {\bf i, ii, iii, iv}. Por komparado, la streklinio rilatas al preterpaso (\ref{Dop}), kun $b\!=\!a/10$.}                 \label{Figuro4} 
\end{figure}

Por skribi $\nu_{obs}/\nu$ kiel funkcio de nur $x$ kaj $\epsilon$, ni uzas (\ref{t}) en (\ref{Dop0}),   
\bea                                                  \label{Dop0x}
\nu_{obs}/\nu\!=\!1+a+x-\epsilon\epsilon_x\sqrt{(a+1+x)^2-1}\ . 
\eea 
Figuro~\ref{Figuro4} montras la malkontinuecon de $\nu_{obs}/\nu$ je la momentoj $-t_P$ kaj $t_P$. La valoro de tiu malkontinueco estas, de  (\ref{Dnova1}), $2t_P$.  

\section{Observanta propratempo} 
La observanto povas preferi uzi sian propratempon $\tau$, anstata^u uzi la tempon $t$ de la fonta horlo^garo. La taktoj de tiuj du tempoj rilatas kiel $\dd\tau\!=\!\sqrt{1-v^2/c^2}\,\dd t$; integralante uzante la rapidon $v$ de~(\ref{vt}) kaj la kondi^con $\tau\!=\!0$ kiam $t\!=\!0$, tiu esti^gas 
\bea                                                    \label{ttr}
t\!=\sinh \tau\ . 
\eea 
Uzante $\tau$, la pozicio~(\ref{xt}) kaj la rapido~(\ref{vt}) de la observanto skribi^gas 
\bea                                                    \label{xtr}
x\!=\!-a+\cosh \tau-1, \hskip3mm  v\!=\!\tanh \tau\ ,
\eea 
kaj la momentoj~(\ref{tP}) de preter(tra)paso esti^gas $\mp \,\tau_P$, estante 
\bea                                                    \label{tr1}
\tau_P\!=\cosh^{-1}(1+a)\ .
\eea  

\begin{figure}[ht]                                             
\centerline{\epsfig{file=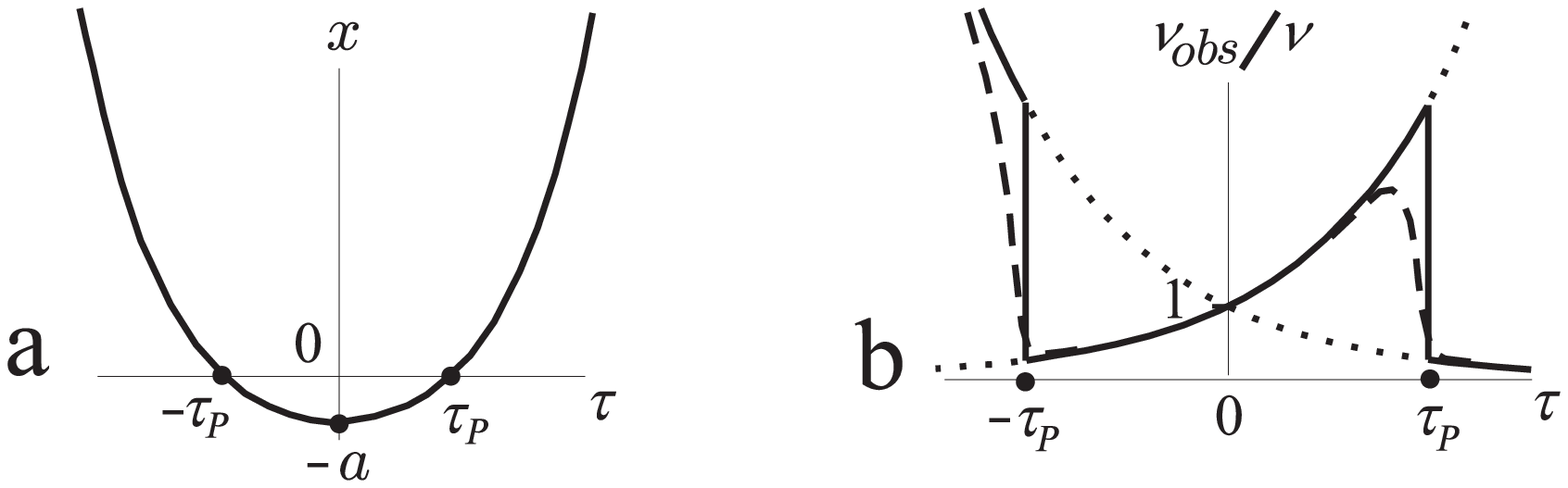,width=12cm}} 
\caption{
\newline 
{\bf \ref{Figuro5}.a} Katenario montranta pozicion $x$ kiel funkcio (\ref{xtr}) de observanta propratempo $\tau$.  
\newline  
{\bf \ref{Figuro5}.b} montras $\nu_{obs}/\nu$ kiel funkcio de observanta propratempo $\tau$. La plenlinio rilatas al trapaso $(b\!=\!0)$, kaj la streklinio rilatas al preterpaso (\ref{Doplin}) kun $b\!=\!a/10$. Punktlinioj montras eksponencialojn e$^{-\tau}$ kaj e$^{\tau}$.} 
                                                         \label{Figuro5}
\end{figure}

^Ce preterpaso $(b\!\neq\!0)$ la Dopplera efekto (\ref{Dop}) kiel funkcio de $\tau$ skribi^gas 
\bea                                                 \label{Doplin}
\nu_{obs}/\nu\!=\!\cosh \tau-\frac{x\,\sinh \tau}{\sqrt{x^2+b^2}}\ ,  
\eea 
kie $x(\tau)$ estas ^ce~(\ref{xtr})\ . Atentu ke ^ce trapaso ($b\!=\!0$), la (\ref{Doplin}) multe simpli^gas: 
\bea                                                  \label{seila}
\nu_{obs}/\nu =
\left\{
\begin{array}{l}
{\rm e}^{-\tau}, \hskip2mm |\tau|\!>\!\tau_P\ ,
\\  
{\rm e}^{\tau}\ , \hskip3mm |\tau|\!<\!\tau_P\ . 
\end{array}
\right.
\eea 
Figuro~\ref{Figuro5}.b montras kontinuan funkcion (\ref{Doplin}) kaj nekontinuan funkcion (\ref{seila}). 

\section{Komentoj}
La grafiko 3.a montras onduman formon; tio okazas, ^car la ekvacio~(\ref{semnome}) de  Dopplera efekto havas du faktorojn. La unua faktoro dependas kvadrate je la rapido $v$, kaj ka^uzas ^ciam viol-delokigon. La dua faktoro dependas lineare je la rapido, kaj dependas anka^u je la pozicio $x$\ ; ^ci tiu faktoro povas ka^uzi ru^g- a^u viol-delokigon. Estas la batalo inter tiuj du faktoroj kiu determinas la onduman formon de la Dopplera efekto. 

Nova artikolo estas skribonta, {\em Relativeca Dopplera efekto ^ce unuforme akcelata movo -- II}. Tie ni studos okazojn kies la observanto restas, kaj la eliganta fonto movi^gas pasante preter a^u tra la observanto.  

\begin{figure}[ht]                                             
\centerline{\epsfig{file=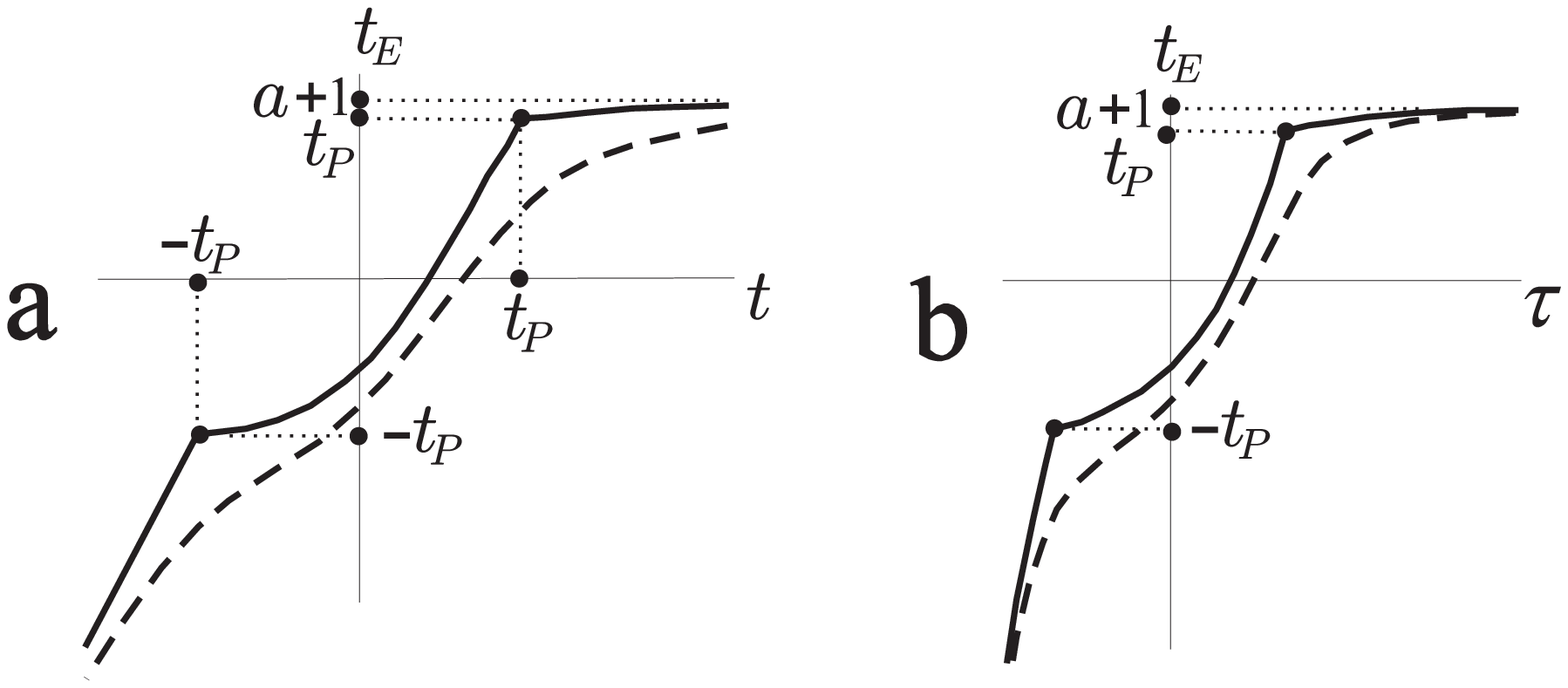,width=10cm}} 
\caption{
\newline 
{\bf \ref{Figuro6}.a} Funkcio $t_E(t)$, 
ekvacio (\ref{tE}). La plenlinio rilatas al trapaso $(b\!=\!0)$, kaj la streklinio rilatas al preterpaso kun $b=a$. Vidu asimptoton $t_E=a+1$.  
\newline  
{\bf \ref{Figuro6}.b} Funkcio $t_E(\tau)$, ekvacioj (\ref{ttr}) kaj (\ref{tE}). La plenlinio rilatas al trapaso $(b\!=\!0)$, kaj la streklinio rilatas al preterpaso kun $b=a$. Vidu asimptoton $t_E=a+1$.} 
                                                         \label{Figuro6}
\end{figure}

Fine, ^sajnas interesa vidu kiel interrilati^gas, la pluraj tempoj uzataj en ^ci tiu artikolo. Figuro~\ref{Figuro6} montras $t_E(t)$ kaj $t_E(\tau)$, por kaj preterpaso kaj trapaso. Vidu ^ce \ref{Figuro6}.b ke la inklinacioj $\dd t_E/\dd\tau$ estas la Dopplera faktoro $\nu_{obs}/\nu$, kiel ^ce figuro~\ref{Figuro5}.b.

\end{document}